\input epsf
\magnification=\magstep1
\noindent
\def \half {{\scriptstyle {1\over 2}}}

\centerline{\bf Phase Transitions}
\medskip
\centerline{Michael Creutz}
\centerline {Physics Department}
\centerline {Brookhaven National Laboratory}
\centerline {Upton, NY 11973}
\medskip
\centerline {creutz@bnl.gov}
\bigskip

This is a set of notes on phase transitions and critical phenomena
prepared to accompany my lectures for the RHIC '97 summer school, held
at Brookhaven from July 6 to 16, 1997.

\medskip
\medskip
\hrule
\bigskip

I have been asked to provide an elementary introduction to phase
transitions and critical phenomena.  The subject is vast; thus, this
can really only be a somewhat superficial personal overview of the
subject.  Many important topics are left out; at the end of these
notes is a brief bibliography of sources for further reading.

While most of the universe is a dilute gas of hydrogen, phase
transitions are crucial to our existence.  We breath air, blood flows
through our veins, and our bones are solid.  Water boils and lakes
freeze over.  Mathematically, however, phase transitions are rather
remarkable.  Statistical mechanics is based on Boltzmann weights
$$
P(S)={e^{-\beta E(S)}\over Z}
$$
where the inverse temperature $\beta={1\over kT}$.  From these the
partition function
$$
Z=\sum_S e^{-\beta E(S)}
$$
is constructed.  But $e^{-\beta E}$ is an analytic function of
$\beta$, {\it i.e.} it has no singularities.

Phase transitions can only occur when an infinite number of states are
available, such as with infinite volume $V$.  Start with the partition
function
$$
Z=\sum_E N(E) e^{-\beta E}
$$
where $N(E)$ is the number of states of a given energy $E$.  Each
piece of a large system can have its own energy, so one should expect
$\langle E \rangle$ proportional to $V$.  Work with the energy density
${\cal E}=E/V$.  Since there are of order $V$ places to put each bit
of energy, we also expect the number of states of a given energy
density to grow exponentially
$$
N(E)=\exp(VS({\cal E}))
$$
This defines the entropy density $S$ at the given energy density.
Pulling out the volume factors explicitly,
$$
Z=V\int d{\cal E} e^{V(S({\cal E})-\beta{\cal E})}
$$
As $V\rightarrow \infty$ the integral is dominated by the
maximum of the integrand, where
$$
0={d\over d{\cal E}}(S({\cal E})-\beta{\cal E})=
{\partial S\over \partial E}-\beta
$$
or the more usual form
$$
\Delta S=\Delta Q/T
$$
In this saddle point approximation
$$
Z\sim e^{-\beta VF}
$$
where free energy is
$$
F={\cal E}-S/\beta 
$$
and all of thermodynamics follows.

A singularity at a phase transition requires structure in $N(E)$.
Such structure also requires spatial correlations.  Otherwise
$$
Z(V)\sim Z(V/N)^N 
$$  
and the free energy becomes just that for a small volume.

In much of the following I will use the Ising model as an example.
This has ``spins'' $s_i\in\{1,-1\}$ occupying lattice sites and has
energy $E=-\sum_{\{ij\}} s_is_j$, where $\{ij\}$ denotes nearest
neighbors.   Table 1 gives $N(E)$ for the two dimensional Ising
model on 10 by 10 lattice with periodic x boundaries, cold y walls.
The length of the line gives the entropy.  The phase transition is
hidden in a subtle flatness.

The numbers in this table are rather large.  They add to
$2^{100}=1.267\times 10^{30}$.  Our universe is only about $10^{27}$
nanoseconds old, suggesting that it is impossible to calculate this
list by simple counting.  This is a frequent argument for Monte Carlo.
See my state-counting papers to learn how I got this list.

\topinsert
\centerline{Table 1. State counts for the Ising model on a 10 by 10 lattice.}
\medskip
\noindent $E/2+100$ \qquad $N(E)$ \hfil $E/2+100$ \qquad $N(E)$ \hfill
\hrule
\smallskip
{\sevenrm 
\baselineskip=6pt
\obeylines
\hbox{
\vbox{\hsize=.5\hsize
\obeylines
0 \qquad   1
2 \qquad   0
4 \qquad   100
6 \qquad   190
8 \qquad   5390
10 \qquad   19920
12 \qquad   226185
14 \qquad   1123330
16 \qquad   8441545
18 \qquad   46439270
20 \qquad   288232165
22 \qquad   1596503840
24 \qquad   9008597170
26 \qquad   48530806690
28 \qquad   258919598835
30 \qquad   1348085135068
32 \qquad   6918375532625
34 \qquad   34921952998720
36 \qquad   173864285141465
38 \qquad   853528946161100
40 \qquad   4131702217991006
42 \qquad   19698116107747500
44 \qquad   92337394182797240
46 \qquad   424635096183933970
48 \qquad   1910993686546702565
50 \qquad   8394325581182421100
52 \qquad   35900636024138056610
54 \qquad   149134699701274540190
56 \qquad   600434187444808042305
58 \qquad   2338237484656296289710
60 \qquad   8790991827530812668453
62 \qquad   31852806882802872810510
64 \qquad   111039862678342970767760
66 \qquad   371793726574328382611580
68 \qquad   1193670523583033542771745
70 \qquad   3668437423804485582262430
72 \qquad   10772807184138254585743365
74 \qquad   30174747119602748554894980
76 \qquad   80467250627920555722255415
78 \qquad   203904785227407787528278180
80 \qquad   490026517327332203099130689
82 \qquad   1114622254786255520262613920
84 \qquad   2394787743912498152267010800
86 \qquad   4849969799910449080522379200
88 \qquad   9239228193366464342451697155
90 \qquad   16521328755364544210468233924
92 \qquad   27673114057688890670065067455
94 \qquad   43328960149817735987320787580
96 \qquad   63289600282274727602148469440
98 \qquad   86076254527328476831763676120
}
\vbox{\hsize=.5\hsize
\obeylines
100 \qquad   108804232426376087683496097815
102 \qquad   127615138775266749696010320050
104 \qquad   138682226083589753382353631155
106 \qquad   139467535997338317070747513220
108 \qquad   129673265537564086898449474485
110 \qquad   111398087687361442602934363604
112 \qquad   88394194656000609637107306835
114 \qquad   64789735278060885125545778420
116 \qquad   43882526876091406802688842040
118 \qquad   27484620182084875609413209920
120 \qquad   15934677821408488316923097025
122 \qquad   8562769731912107647661352420
124 \qquad   4271377195758556988860024315
126 \qquad   1981325557749996784540426400
128 \qquad   856247175668720270761391365
130 \qquad   345440085480687517414714344
132 \qquad   130373941135805243213306725
134 \qquad   46131131663242989983156880
136 \qquad   15336949736067657882440975
138 \qquad   4801625511818556981759340
140 \qquad   1418746354667950902604900
142 \qquad   396504230728933768862650
144 \qquad   105044804469611713155910
146 \qquad   26439076355718752657610
148 \qquad   6336377505749490029695
150 \qquad   1449347253869825330984
152 \qquad   317184792213120157975
154 \qquad   66590745159525686410
156 \qquad   13450173814318534170
158 \qquad   2621824707749641960
160 \qquad   494837291835094171
162 \qquad   90726699739843320
164 \qquad   16209249292505960
166 \qquad   2829255985524290
168 \qquad   483344637121035
170 \qquad   80889449574800
172 \qquad   13259776474415
174 \qquad   2126884521530
176 \qquad   333319272600
178 \qquad   50912615760
180 \qquad   7565408818
182 \qquad   1088231770
184 \qquad   151489010
186 \qquad   20119550
188 \qquad   2579540
190 \qquad   303762
192 \qquad   35230
194 \qquad   3340
196 \qquad   350
198 \qquad   20
200 \qquad   2
}
\hfill
}

\par
} 
\endinsert

\bigskip
How can we get enough correlation for a phase transition?  One way is
surface tension.  Let small volumes be in two possible phases, {\it
i.e.} water and steam.  Suppose we pay a penalty for an interface
between the phases.  For a model, put the system on a lattice with two
states in each cell, $s_i\in \{1,-1\}$, representing water and steam
$$
Z= \sum_{s_i} e^{-\beta \sum_i F(s_i) +\sum_{nn}J s_i s_j}.
$$
Here $J$ represents the surface tension and $nn$ means nearest
neighbors.  The relative free energies of the two states can be
influenced by, say, the pressure.  The interesting case is when they
are near each other, so let me expand
$$
F(s_i)= \overline F-H s_i
$$
where $\overline F=(F(1)+F(-1))/2$ and $H=-(F(1)-F(-1))/2$.

\topinsert
\epsfxsize .5\hsize
\centerline{\epsffile {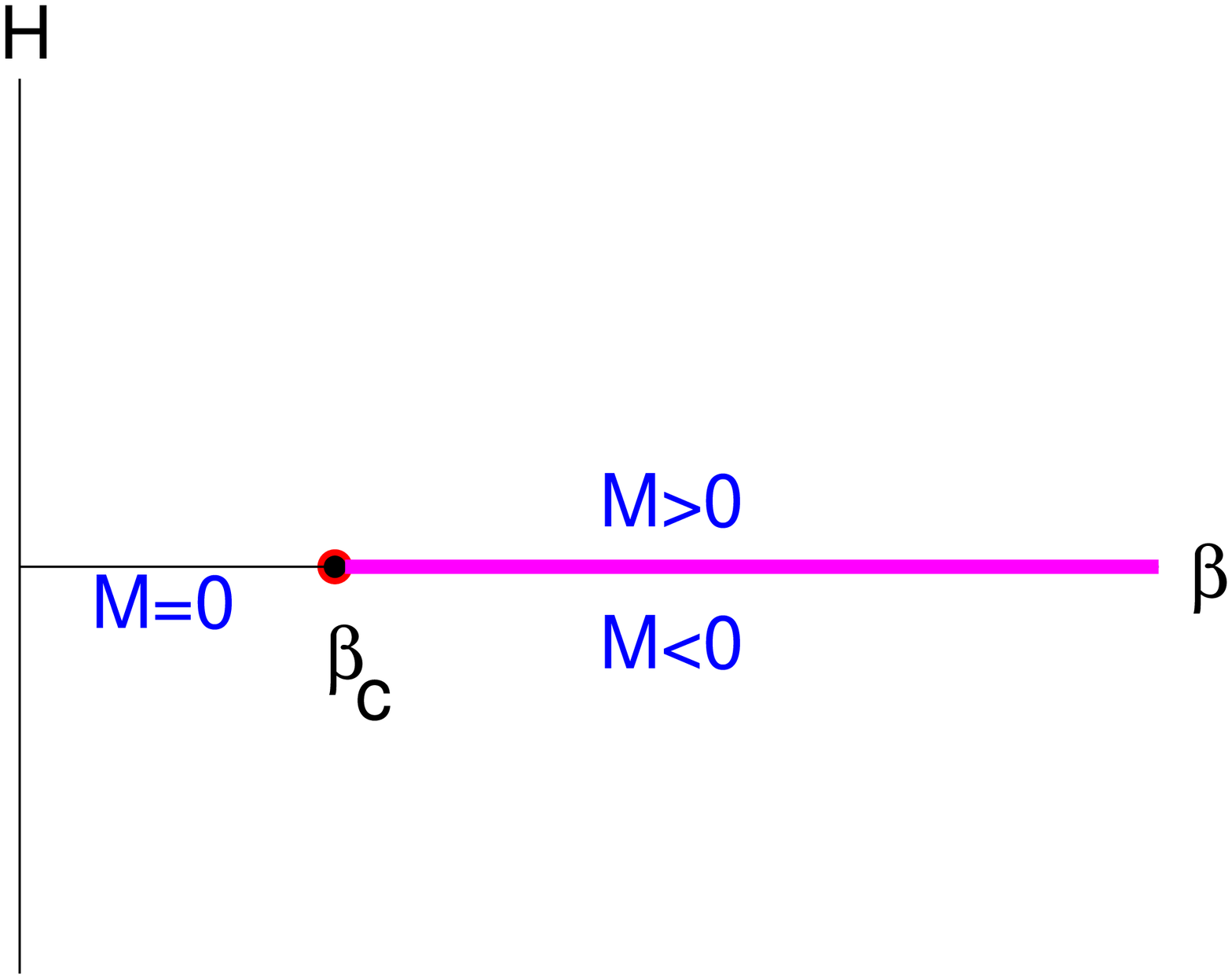}}
\centerline{Figure 1. Phase diagram for the Ising model.}
\endinsert

Thus we expect the boiling of water to be described by something like
the Ising model in an applied field, more usually used for magnets
$$
Z= \sum_{s_i} e^{\beta \sum_{nn} s_i s_j + H \sum_i s_i}
$$
Boiling occurs on a flip in the sign of $H$ at low temperature.  The
basic structure of this model, shown in Fig.~1, is a first order line
running from a critical $\beta$ to infinity along the $\beta$ axis.
The $H\leftrightarrow -H$ symmetry, expected for magnets, is broken in
water by higher order effects.

\topinsert
\epsfxsize .5\hsize
\centerline{\epsffile {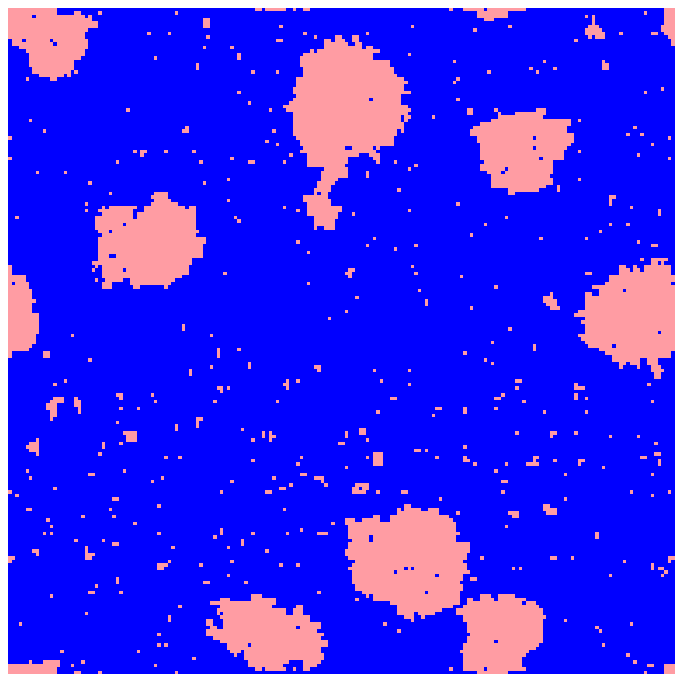}}
\centerline{Figure 2. Simulation of boiling water.}
\endinsert

Fig.~2 shows a simulation of water boiling, obtained by taking an
ordered Ising system and turning on a field in the opposite direction
to the magnetization.  The bubbles nucleate from larger fluctuations.
The picture was made using xpotts, a Potts model simulator from my
xtoys collection at http://penguin.phy.bnl.gov/www/xtoys/xtoys.html.

\bigskip
OK, so we need infinite volume, but how infinite?  Try a chain of
sites in a ring, let the ring length go to infinity.  A bond in the
chain connecting $s$ and $s^\prime$contributes
$$
T_{s,s^\prime}=e^{\beta s s^\prime + H (s+s^\prime)/2}
=\pmatrix{
e^{\beta+H} & e^{-\beta}\cr
e^{-\beta} & e^{\beta-H}\cr
}_{s,s^\prime}
$$
This is the ``transfer matrix.''  Summing over spins gives the
partition function for an $N$ site lattice
$$
Z={\rm Tr} (T^N)
$$
This can be calculated by diagonalizing $T$
$$
Z=\lambda_+^N+\lambda_-^N
$$
with
$$
\lambda_\pm=e^\beta\left(\cosh(H)\pm\sqrt{\sinh^2(H)+e^{-4\beta}}\right)
$$
The free energy per site is dominated by the largest eigenvalue
$$
-\beta F={\log Z \over N}=\log(\lambda_+) 
+ \exp(-N \log(\lambda_+/\lambda_-)) +\ldots
$$
Since $\lambda_+$ is analytic and positive, the infinite volume free
energy has no singularities.  We fail to find a phase transition.
Note that the finite volume corrections are exponentially small.  This
shows that the theory has a mass gap.  As $\beta$ goes to infinity
with $H=0$, the eigenvalues become equal and the mass gap goes to
zero.  The phase transition, to the extent there is one, occurs at
zero temperature.

Physically, a kink anti-kink pair has a finite probability, but an
infinite number of possible separations.  This infinity always
disorders the system.  In more dimensions a big bubble pays a penalty
proportional to its surface, so it is suppressed.  For now I will
continue to concentrate on the $H=0$ Ising model, but with more
neighbors.

\topinsert
\epsfxsize .6\hsize
\centerline{\epsffile {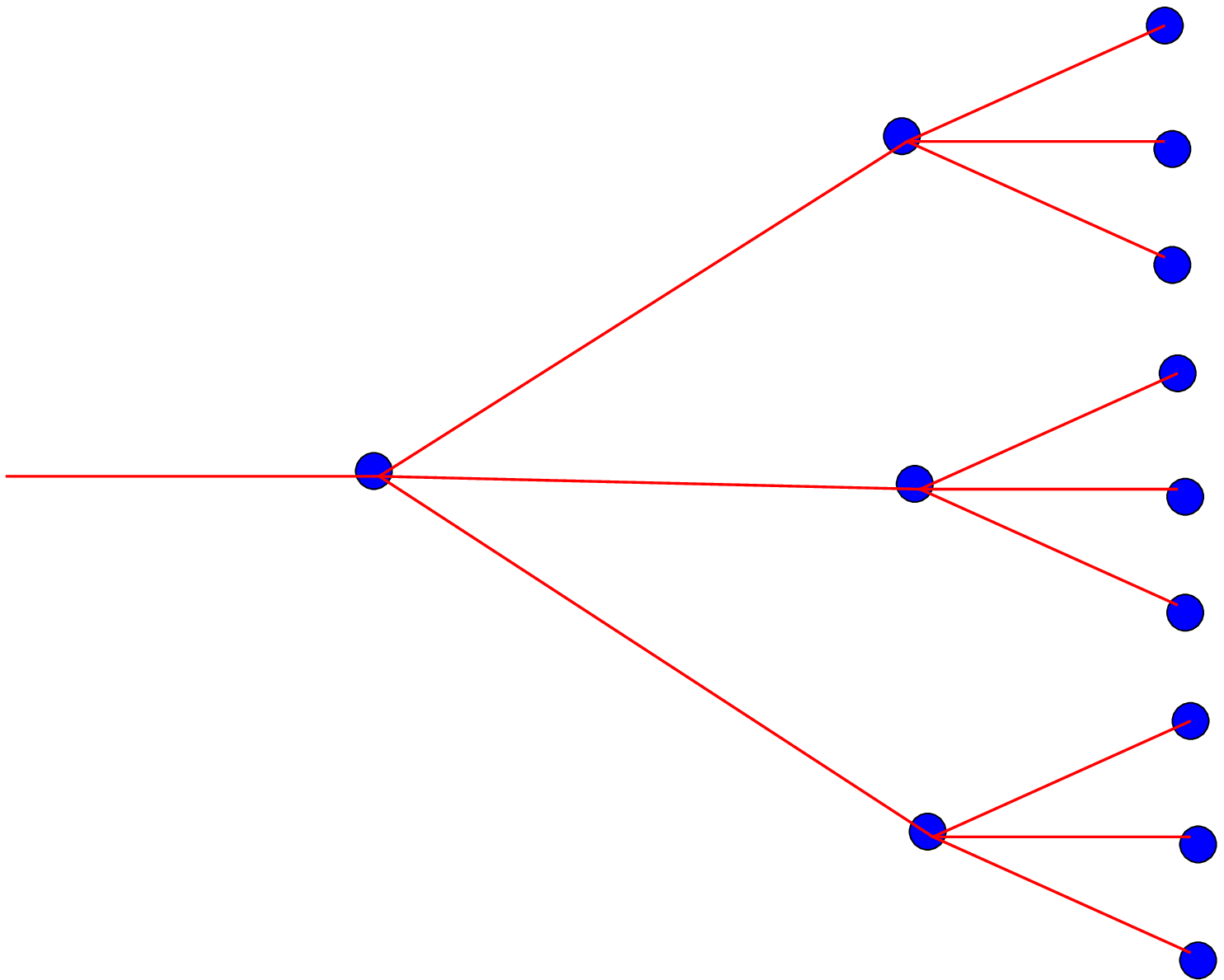}}
\centerline{Figure 3. Constructing a Cayley tree.}
\endinsert

So we need more neighbors.  Now for an amusing model.  Suppose each
site has $2d$ neighbors, but on a system without closed loops.  Let me
build this lattice recursively.  A single outermost site is connected
to a deeper neighbor, which has 2d-1 other neighbors.  This site can
have either spin -1 or spin 1.  Fix this spin and define $Z(s)$ to be
the partition function obtained by summing over all deeper spins.
This partial tree is then recursively built up, defining a Cayley tree
as sketched in Fig.~3.  To get things started, apply an infinitesimal
field to the zeroth level sites
$$
Z_0(s)=e^{\epsilon s}
$$
At level $n$, the partition function can be written as a sum over the
values of the level $n-1$ neighbors
$$\matrix{
Z_n(1)=\left(e^\beta Z_{n-1}(1)+e^{-\beta}Z_{n-1}(-1)\right)^{2d-1}\cr
Z_n(-1)=\left(e^\beta Z_{n-1}(-1)+e^{-\beta}Z_{n-1}(1)\right)^{2d-1}\cr
}
$$
Dividing we obtain
$$
R_n\equiv{Z_n(1)\over Z_n(-1)}=\left({e^\beta R_{n-1}+e^{-\beta}\over 
           e^\beta+e^{-\beta}R_{n-1} }\right)^{2d-1}
$$

\topinsert
\epsfxsize .6\hsize
\centerline{\epsffile{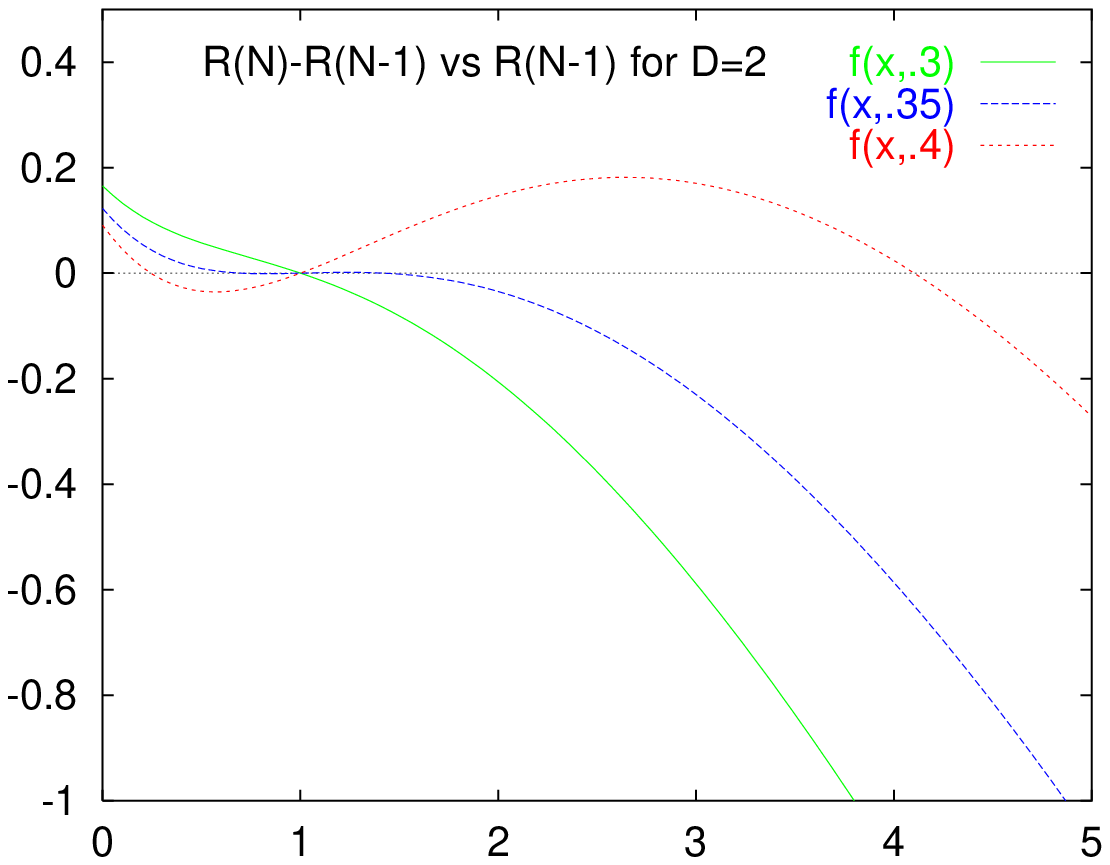}}
\centerline{Figure 4.  Recursive solution of the Ising model on a Cayley tree.}
\endinsert

We can now look for a stable asymptotic solution for a nontrivial $R$
by asking that $R_n=R_{n-1}=R$.  As shown in Fig.~4, $R=1$ is always
one solution, but more appear at the critical coupling which occurs
when the derivative of the right hand side is unity.  This happens at
$$
\tanh(\beta_c)={1\over 2d-1}
$$
When $\beta$ exceeds the critical value, a value of $R$ above one will
flow towards the non-trivial fixed point as seen in the figure.  The
transition for this model is second order since the fixed point
smoothly moves to unity as the critical coupling is approached.  This
concept of iteration generating a ``flow'' towards a ``fixed point''
will be a recurring theme in the following.  Note that as $d$ goes to
one the critical temperature moves to infinity.  One dimension is a
critical case.

\bigskip
I now digress into the topic of duality, which gives the exact
critical point for the 2d case.  To see how this works, change
variables from sites to bonds.  This can be done a couple of different
ways.

For a given configuration, each bond is either excited or not.  An
excited bond gives a factor of $e^{-\beta}$.  A non-excited bond gives
$e^{\beta}$.  Re-express the sum over sites as a sum over bonds being
excited or not.  These variables over-determine the spins, and thus are
constrained; in going around a loop one must encounter an even number
of excited bonds.  The product of bonds around a plaquette is
positive.  This view of the system forms the basis for the low
temperature expansion, {\it i.e.} at low temperature most bonds not
excited.

Now for an alternative view, write $e^{\beta s_i
s_j}=\cosh(\beta)+s_is_j \sinh(\beta)$.  On each bond either take one
or the other of these two terms.  Assign 1 to bonds with the first
choice, $-1$ to the others.  Given a particular configuration of such
choices, sum out the spins.  This will give zero unless there are an
even number of bonds coming out of a site which use the $\sinh(\beta)$
term.  This expansion on the bonds is the basis of the high
temperature expansion, {\it i.e.} small $\beta$ means small
$\sinh(\beta)$.  Again we get a sum over two states for each bond,
with a constraint on these choices.  The bond variables multiplied
over the four links coming from any bond must be positive.

\topinsert
\epsfxsize .4\hsize
\centerline{\epsffile{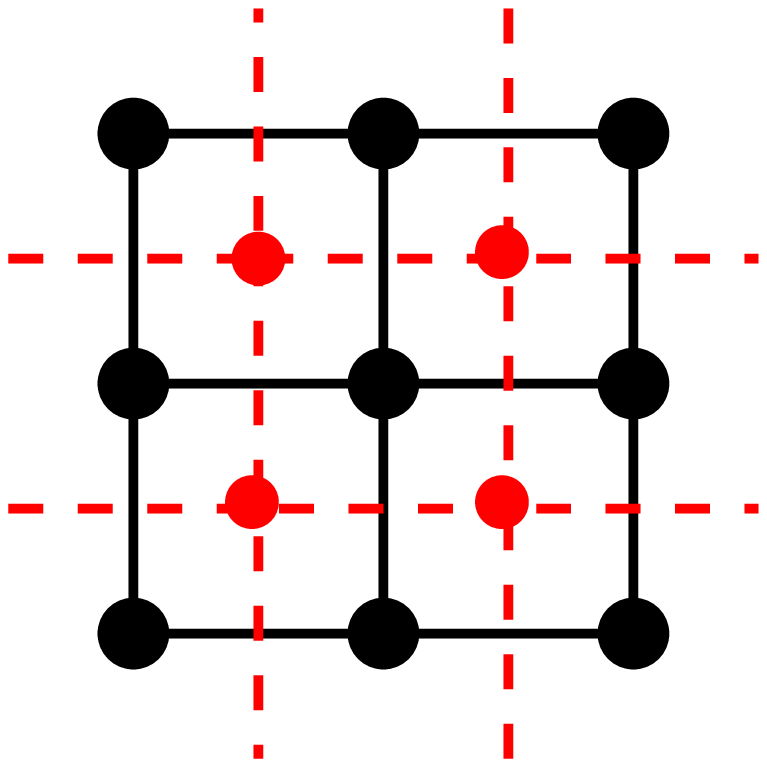}}
\centerline{Figure 5. A two dimensional lattice and its dual.}
\endinsert

These two constrained systems are closely related.  Transfer the
variables from the bonds to dual ones, new bonds crossing the middle
of each old one, new sites in the middle of the old plaquettes, as
shown in Fig.~5.  The product of old bonds out of a site becomes a
product of new bonds about the new plaquette.  With this mapping on
one of the bond descriptions, the two constrained systems now satisfy
the same constraint.  This means that up to irrelevant factors, the
physics at two values of $\beta$ is related
$$
e^{-2\beta}=\tanh(\beta^\prime)
$$
Under this mapping strong and weak limits interchange.  If there is
only one singularity, it must occur at the self dual point
$\beta=\beta^\prime=\beta_c ={1\over 2} \log(1+\sqrt 2)=.44068\ldots$.
This idea of duality extends to $Z_n$ clock models, Potts models, 4-d
$Z_2$ lattice gauge theory, 3d gauge-Higgs system, QED with monopoles,
$\ldots$ Seiberg $\ldots$.  But it all starts here.

\bigskip
I now turn to mean field theory.  Consider a large number of
dimensions.  Then each site has lots of neighbors, suggesting we might
assume their effect can be averaged.  Suppose we are in a magnetized
state with $\langle s \rangle=M$.  Look at one spin in the average
field of the others.  Calculate its magnetization
$$
M={e^{2dM\beta}-e^{-2dM\beta}\over e^{2dM\beta}+e^{-2dM\beta}}
=\tanh(2dM\beta)
$$
This can be solved by iterating
$$\eqalign{
&M\rightarrow M+\Delta M\cr
&\Delta M=\tanh(2dM\beta)-M\cr
}
$$

\topinsert
\epsfxsize=.6\hsize
\centerline{\epsffile {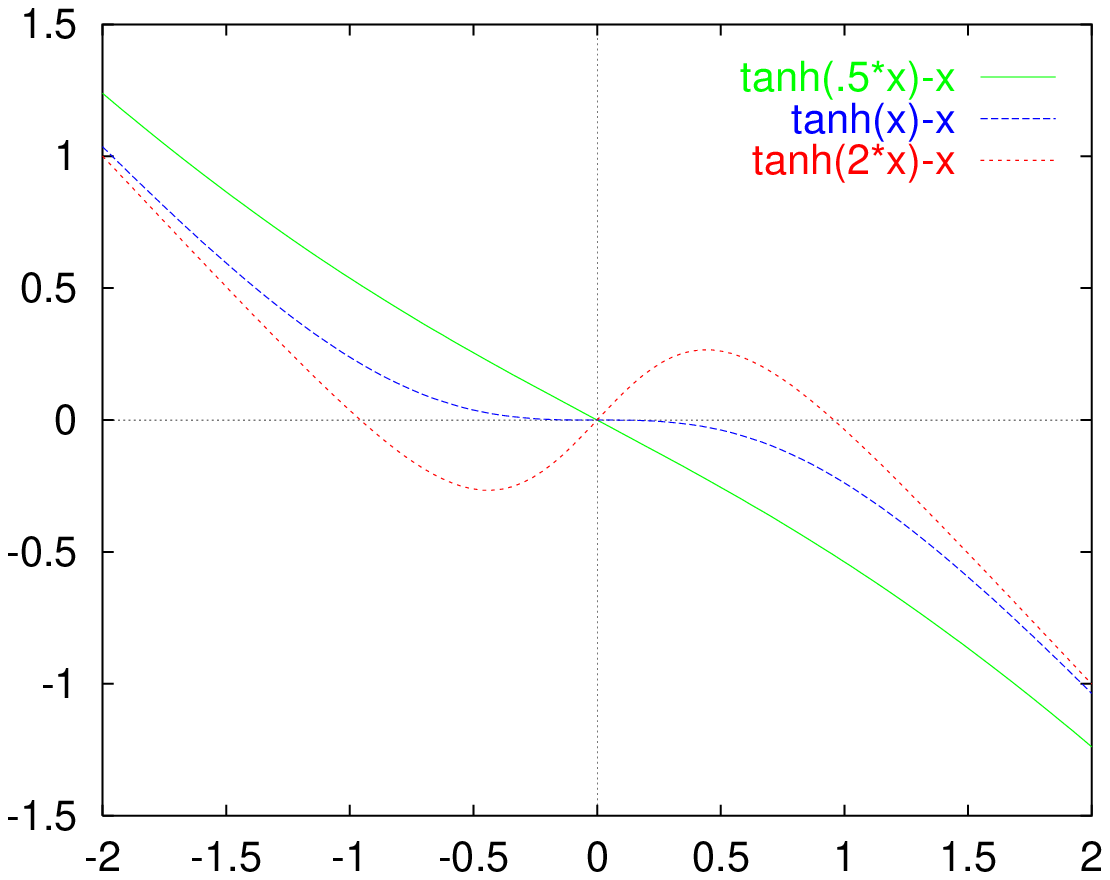}}
\centerline{Figure 6. Recursively solving the mean field equation.}
\endinsert

\noindent We have a non-trivial fixed point only if 
$$
\beta\ge \beta_c={1\over 2d}
$$
As shown in Fig.~6, this ``flow'' like picture gives a similar result
to the Cayley tree case at large d, but the approximation misses the
transition moving to infinity as the dimension goes to one.

\bigskip
Jensen's inequality provides a more formal approach to mean field
theory. Note that $e^x$ is a convex function, ${d^2\over
dx^2}e^x=e^x\ge 0$.  If $x$ is some stochastic variable, this means
that $\langle e^x \rangle > e^{\langle x \rangle}$.  Show this for
homework.

Consider a ``fake'' weighting with $P(s_i)={e^{Hs_i}\over
e^{H}+e^{-H}}$.  With this probability distribution
$$
\langle s_i \rangle_P=\tanh(H)
$$
Thus $H$ might be thought of as a ``source'' pulling on the spins.
Now I manipulate the partition function
$$\eqalign{
Z&=\sum_{\{s\}} e^{\beta \sum_{\{ij\}}s_is_j}\cr
&=\sum_{\{s\}} e^{\beta \sum_{\{ij\}}s_is_j-\sum_i \log(P(s_i))}
\prod_i P(s_i)\cr
&=\langle e^{\beta \sum_{\{ij\}}s_is_j-H\sum_i s_i+V\log(2\cosh(H))} 
\rangle_P\cr
&\ge \exp(\langle {\beta \sum_{\{ij\}}s_is_j-H\sum_i s_i+V\log(2\cosh(H))} 
\rangle_P)\cr
&=\exp(V(d\beta \tanh^2(H)-H\tanh(H)+\log(2\cosh(H))))\cr
}
$$
Thus for any $H$ the true free energy is less than
$$
F\le=F_{mf}=-d\beta \tanh^2(H)+H\tanh(H)-\log(2\cosh(H)).
$$

\topinsert
\epsfxsize .6\hsize
\centerline{\epsffile {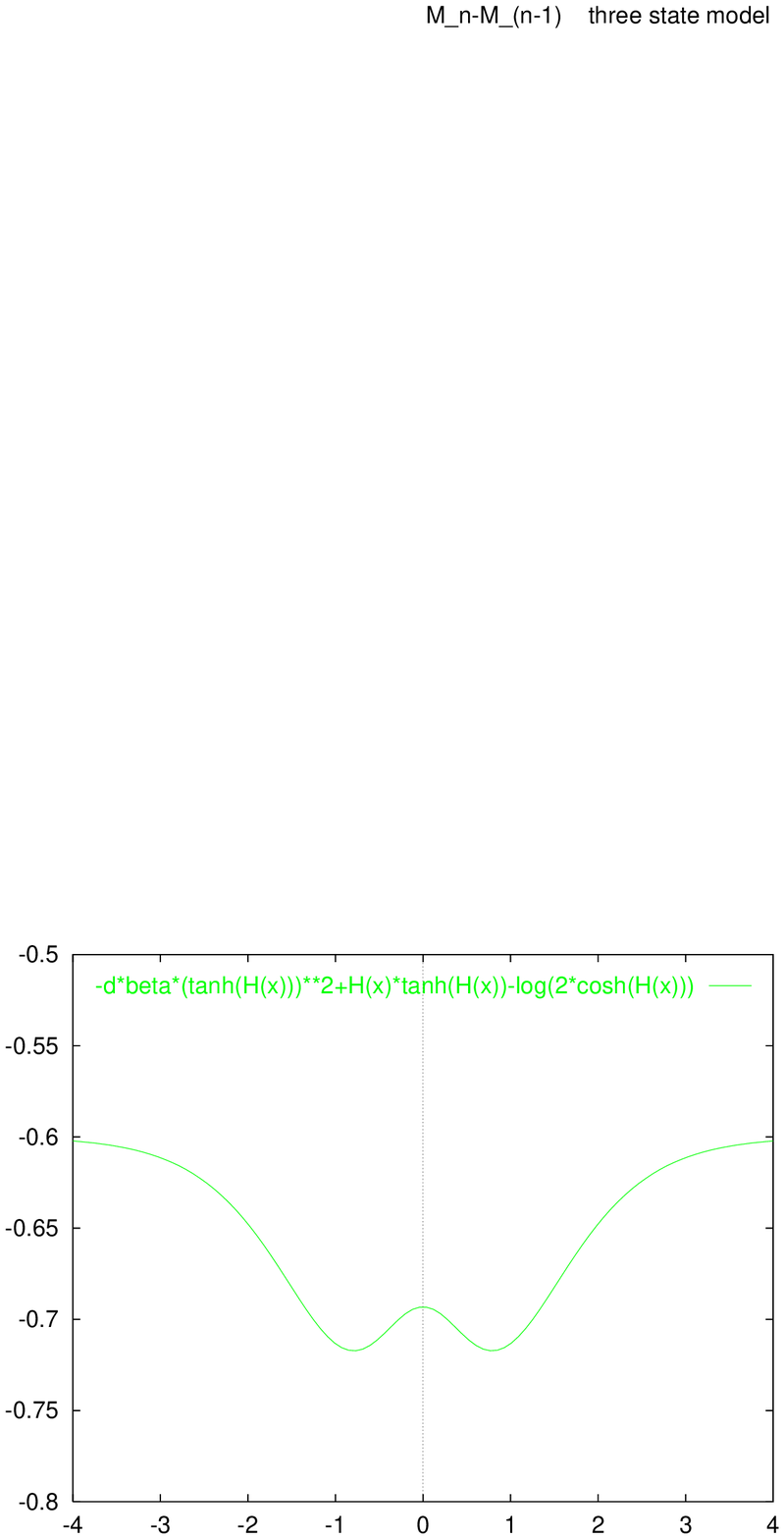}}
\centerline{Figure 7.  The effective potential from Jensen's inequality.}
\endinsert

This is an ``effective potential'' which, depending on the value of
$\beta$, can have one minimum at $H=0$ or a double well shape with two
minima, as sketched in Fig.~7.  The latter represents the ordered
phase.  The critical point occurs when ${\rm O}(H^2)$ term vanishes,
{\it i.e.} the ``mass term.''  This happens at
$$
d\beta_c-1/2=0
$$ 
or $\beta_c={1\over 2d}$, as before.

\bigskip
These transitions have been second order, {\it i.e.} the system
evolves continuously with the coupling parameters.  The picture is a
bit different with three states, where cubic terms can drive us to
first order transitions, and physics becomes discontinuous.  For
example consider the three state Potts model, a system with a $Z_3$
symmetry.  Take $s_i\in\{1,e^{2\pi i/3},e^{-2\pi i/3}\}$.  Suppose the
bonds have low energy if the spins are ``parallel'' or equal, higher
energy otherwise
$$
E=-\sum_{ij}{\rm Re}s_i^* s_j
$$
Define the magnetization to vanish for a random distribution
$$
M=\langle {\rm Re}s_i \rangle
$$
For mean field theory, we are to replace neighbors with the average
and solve self consistently.  Now there are two anti-parallel cases,
each giving minus half a unit of magnetization
$$
M={e^{d\beta M}-e^{-d\beta M/2}\over e^{d\beta M}+2e^{-d\beta M/2}}
$$
Expand the right hand side for small $M$
$$
M={d\over 2}\beta M +{d\over 8}\beta^2 M^2 + {\rm O}(M^3)
$$

\topinsert
\epsfxsize .6 \hsize
\centerline{\epsffile{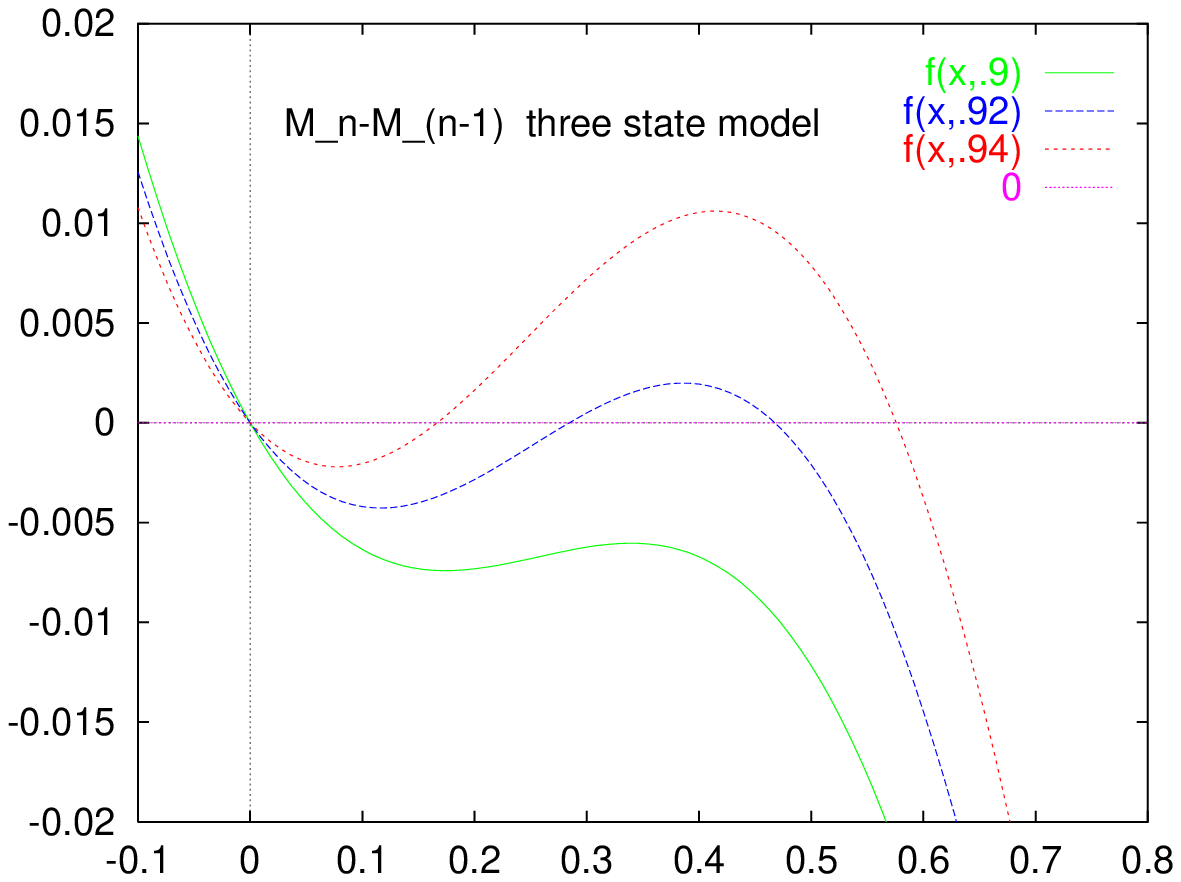}}
\centerline{Figure 8.  Mean field theory for the three state Potts model.}
\endinsert

For the Ising case there was no $O(M^2)$ piece, and the $O(M^3)$ piece
was negative.  As shown in Fig.~8, now a non-trivial solution appears
before reaching $\beta=2/d$.  The new solution appears
discontinuously; when it becomes of lower energy, we have a first
order transition!  Note that when the extra solution first appears, it
is meta-stable and not the lowest energy; one can use the Jensen
inequality argument to estimate when it drops below the unmagnetized
state.  Simulation results show that for three and more dimensions the
transition is indeed first order.  In two dimensions, however, it is
second order and the prediction fails.  In one dimension there is no
transition, just as for the Ising case.

\bigskip

Lattice gauge theory is summarized by the path integral
$$
Z=\sum_U \exp(\beta\sum_P {\rm ReTr}U_P)
$$
with $U_P=U_1U_2U_3U_4$, and the $U_i$ are link variables running
around the plaquette in question.  The local symmetry
$U_{ij}\rightarrow g_i U_{ij} g_j$ implies there is no barrier to
twirling a local group of links around.  Without gauge fixing $U$
cannot have an expectation value for any $\beta$.  This is Elitzur's
theorem.  One should play the Jensen game for more rigor, but proceed
naively anyway, trying to find a self consistent expectation for a
link.  Do $Z_2$ for simplicity, which gives the same result as the
above Ising case except for the replacement $M\rightarrow M^3$ for the
average field
$$
M=\tanh(\beta M^3)
$$
Now there is no linear term at all on the right hand side.  The
prediction is for a strong first order transition.  Most lattice gauge
transitions in fact are first order: $Z_{2-4}$ in four dimensions; all
known gauge groups in 5 or more dimensions.  However, in 3-d, the
$Z_2$ gauge model is dual to the Ising model; so, the transition is
second order.  In 2-d, gauge fixing turns a gauge model into a one
dimensional spin system, with no transitions for any finite
dimensional group.

Later in these lectures I will generalize these arguments to suggest a
first order deconfining transitions for pure $SU(3)$ gauge theory at
finite temperature and for the chiral transition with three massless
quarks.  This will involve some mathematical formalism that I
postpone.

\bigskip
Now I change the subject a bit and remind you of the formal connection
between path integrals and statistical mechanics.  This is one reason
quantum field theorists are also interested in phase transitions.  Let
me start with a simple quantum mechanics problem with the Hamiltonian
$$
H=p^2/2+V(x)
$$
Here $p$ and $q$ are conjugate variables with $[p,q]=-i$.
Look at
$$
Z={\rm Tr} e^{-\beta H}
$$
{\it i.e.} zero space dimensional quantum statistical mechanics.  As
$\beta\rightarrow \infty$ we project out the ground state and get
ordinary quantum mechanics at zero temperature.  This is also the
trace of the evolution operator $e^{-itH}$ for imaginary time
$t=-i\beta$.  Quantum mechanics in imaginary periodic time is the same
problem as quantum statistical mechanics.

I now break up $\beta$ into a large number of ``imaginary time
slices''
$$
Z={\rm Tr} \prod_1^N e^{-\beta H/N}
$$
Insert a complete set of states at each slice
$$
Z=\int dx_1\ldots dx_N \prod_{i=1}^N \langle x_i | e^{-\beta H/N} |
x_i+1\rangle
$$
where $x_{N+1}\equiv x_1$.  Now for large $N$ we can approximate
$$
\langle x_i | e^{-\beta H/N} | x_{i+1}\rangle\sim
e^{-\beta V(x_i)/N}\langle x_i | e^{-\beta p^2/(2N)} | x_{i+1}\rangle
$$
The second factor can be worked out by inserting a complete set of
momentum eigenstates
$$
\langle x_i | e^{-\beta p^2/(2N)} | x_{i+1}\rangle
=\int dp e^{-ip(x_i-x_{i+1})}e^{-\beta p^2/(2N)}
=\sqrt{2\over\pi} e^{-N(x_{i+1}-x_i)^2/(2\beta)}
$$
Thus we have the simple form 
$$
Z=\int dx_1\ldots dx_N e^{- S}
$$
where the ``lattice action'' is simply
$$
S=a \sum_i V(x_i) +\left({x_{i+1}-x_i\over a}\right)^2
$$
and the lattice spacing $a=\beta/N$.  This defines the path integral,
which formally is a classical statistical mechanics problem in one
dimension.  It represents the thermal dynamics of a polymer, with the
$x_i$ denoting the coordinates of the atoms in the chain.  

Going from a single particle to a field theory, $D$ space dimensional
quantum mechanics is equivalent to $d=D+1$ dimensional classical
statistical mechanics.  The infinite ``time'' limit gives the ground
state, while finite imaginary time relates quantum statistical
mechanics to classical statistical mechanics in one more dimension.

Second order phase transitions are essential to continuum limits.
Taking the lattice spacing to zero requires physical correlation
lengths to diverge in lattice units.  The particle physicist's
$e^{-mr}$ corresponds to the statistical mechanic's $e^{-n/\xi}$.
With lattice spacing $a$, we identify $r=na$ and $\xi=m/a$.  A
continuum limit requires $\xi\rightarrow\infty$, which occurs at a
phase transition, most particularly at a second order one.  For lattice gauge theory,
4-d is a borderline case and the transition occurs at $\beta\sim
1/g^2=\infty$, with the approach given by standard asymptotic freedom
arguments.

\bigskip
Now for another jump in subject.  Despite the usual pedagogical
approaches, effective potentials want to be convex.  In field theory
language, consider
$$
Z=\int d\phi e^{-S(\phi)}
$$
Adding in some external sources
$$
Z(J)=\int d\phi e^{-S(\phi)+J\phi}
$$
general correlation functions can be found by differentiating with
respect to $J$.  Here I shorthand notate $J\phi=\int dx J(x)\phi(x)$
in the continuum, or $J\phi=\sum_i J_i\phi_i$ on the lattice.  Think
of $J$ as an external force pulling on the field.  Such a force will
give the field an expectation value
$$
\langle \phi \rangle_J=-{\partial F\over \partial J}
$$
where I define the free energy $F(J)=-\log(Z(J))$.  Now imagine inverting
this to find what force $J(\Phi)$ gives some desired expectation value,
{\it i.e.} solve
$$
\Phi(J)=\langle \phi \rangle_{J(\Phi)}=-{\partial F\over \partial J}
$$
In terms of this formal solution, construct the ``Legendre transform''
$$
V(\Phi)=F(J(\Phi))+\Phi J(\Phi)
$$
and look at
$$
{\partial V \over \partial \Phi}
=-\Phi {\partial J \over \partial \Phi}+J+\Phi{\partial J \over \partial \Phi}
=J
$$
If I turn off the sources, this derivative vanishes.  Thus the minimum
of $V$ tells us the expectation value of the field.  This quantity $V$
is the ``effective potential.''

But now let me confuse you by looking at the second derivative of $V$
$$
{\partial^2 V \over \partial \Phi^2}={\partial J\over \partial \Phi}
$$
Actually, it is easier to look at the inverse
$$
{\partial \Phi \over \partial J}=
-{\partial^2 F \over \partial J^2}
=\langle \phi^2 \rangle - \langle \phi \rangle ^2
=\langle (\phi-\langle\phi\rangle)^2 \rangle \ge 0. 
$$
Thus this second derivative has a single sign!  This shows we are
actually looking for a minimum and not a maximum of $V$, but in
addition it implies that $V$ can only have ONE minimum!

So what is going on?  Are phase transitions impossible?  The more you
pull, the larger the expectation of $\Phi$ should be.  It won't go
back.  Physically, we must do Maxwell's construction.  If we force the
expectation of $\phi$ to lie between two distinct stable phases, the
system phase separates into a mixture of the two.  Note that there is
no large volume limit in the above discussion.  However other
definitions of $V$ can allow a small barrier at finite volume due to
surface tension effects.  A mixed phase must contain interfaces, and
their energy represents a barrier.

\bigskip
First order transitions have a discontinuity in the internal energy,
representing a latent heat.  The barrier in the effective potential
(modulo the above discussion) allows meta-stability and hysteresis.
Water in a clean container can ``bump'' rather unpleasantly.
Explosives last in a meta-stable state for long periods.  Are things
actually analytic as you pass through the transition?  No, there is an
essential singularity that I will now discuss.  The free energies of
the phases match at the transition; suppose I can expand as we go
through it
$$
\Delta F = C (\beta-\beta_c)
$$
Let the surface tension between the two phases be $\sigma$.
Creating a bubble of radius $r$ costs free energy
$$
E(r)={4\pi r^3\over 3}\Delta F+4\pi r^2 \sigma
$$
This has maximum energy at
$$
0=4\pi r^2 \Delta F +8\pi r \sigma
$$
or
$$
r=-2\sigma/\Delta F
$$
At this point the energy of the bubble is
$$
E_{max}={16\pi\sigma^3\over 3(\Delta F)^2}
$$
As we approach the transition, this radius goes to infinity and this
``semi-classical'' argument becomes rigorous.  This energy represents
a barrier to bubble nucleation, which is suppressed by the Boltzmann
weight
$$
P\sim\exp(-\beta E_{max})=\exp(-{C^\prime \over (\beta-\beta_c)^2})
$$
An essential singularity appears in the physics as one passes into a
meta-stable state.  Even though things look very analytic, they are
not.  Since the meta-stable state can decay, this expression
represents an ``imaginary part'' for the free energy of the unstable
phase.

\bigskip
Now for a brief discussion on some aspects of Goldstone Bosons.
Suppose I have a conserved current
$$
\partial_\mu j_\mu=0
$$
so the corresponding charge $Q=\int d^3x j_0(x)$
is a constant
$$
{dQ\over dt}=-i[H,Q]=0.
$$
Suppose, however, that the vacuum is not a singlet under this charge
$$
Q|0\rangle \ne 0
$$
Then there cannot be a mass gap in the theory.  Consider the state
$$
\exp(i\theta\int d^3x j_0(x) e^{-\epsilon x^2}) |0\rangle
$$
where $\epsilon$ is a convenient cutoff and $\theta$ some parameter.
As epsilon goes to zero this state by assumption is not the vacuum,
but the expectation value of the Hamiltonian goes to zero (normalize
so the ground state energy is zero).  Thus ``spontaneously broken
symmetries'' have no mass gap, {\it i.e.} the theory contains states
of arbitrarily low energy.  These are manifested as massless particles
called Goldstone bosons.

Free massless field theory is a marvelous example of all this
where everything can be worked out.  The massless equation of motion
$$
\partial_\mu\partial_\mu \phi=0
$$
can be written in the form
$$
\partial_\mu j_\mu=0
$$
where 
$$
j_\mu=\partial_\mu \phi.
$$
The broken symmetry is the invariance of the Lagrangian 
$L=\int d^4x(\partial_\mu\phi)^2$ under shifts of the field
$$
\phi\rightarrow\phi+c
$$
Note that $j_0=\partial_0\phi=\pi$, the conjugate variable to $\phi$.
One can work out explicitly
$$
\langle 0 | \exp(i\theta\int d^3x j_0(x) e^{-\epsilon x^2/2}) |0\rangle.
$$
but we can save ourselves the work using dimensional analysis.  The
field $\phi$ has dimensions of inverse length, while $j_0$ goes as
inverse length squared.  Thus $\theta$ above has units of inverse
length.  These are the same dimensions as $\epsilon^2$.  Now for a
free theory by Wick's theorem the answer must be Gaussian in $\theta$,
so we conclude that the above overlap must go as
$$
\exp(-C\theta^2/\epsilon^4)
$$ 
where $C$ is some non-vanishing dimensionless number.  This expression
rapidly goes to zero as epsilon becomes small, showing that the vacuum
is indeed not invariant under the symmetry.  As $\epsilon$ goes to
zero, we obtain a new vacuum that is not even in the same Hilbert
space.  Its overlap with any polynomial of fields on the original
vacuum vanishes.

Two dimensions give some interesting twists on this argument.  Now the
scalar field is dimensionless, and the current has dimensions of
inverse length.  Thus theta is dimensionless and we expect
$$
\langle 0 | \exp(i\theta\int dx j_0(x) e^{-\epsilon x^2/2}) |0\rangle
\sim \exp(-C\theta^2)
$$
There can be no $\log(\epsilon)$ since there is nothing to set the
scale.  Thus the vacuum is not invariant, but the symmetry relation
does not give you a fully independent state.  This is clearly a
borderline case, and for an interacting theory the massless particles
can be lost.

This is tied in with the propagator in two dimensions not being a 
distribution.  Put in a small mass cutoff.  Then 
$$
\Delta(x)=\int {d^2 p\over (2\pi)^2} {e^{-ipx}\over p^2+m^2}
$$
gets very singular as $m$ goes to zero.  Consider the simple test
function $e^{-x^2/2}$.  This is infinitely differentiable and well
behaved at infinity.  Now integrate this test function with the free
particle propagator
$$
\int d^2x \Delta(x)e^{-x^2/2}
\sim\int {d^2 p\over (2\pi)^2} {e^{-p^2/2}\over p^2+m^2}
\sim \log(1/m)\Longrightarrow_{m\rightarrow 0}\infty 
$$
Thus $\Delta(x)$ is not a tempered distribution, contrary to the basic
assumptions of quantum field theory.  However Green's functions of
$j_0$ are tempered distributions since they involve derivatives that
kill the divergent part.  In most cases Goldstone bosons are lost in
two dimensions, however, if they are free, as in the above case, they
can exist.  The X-Y model, with spins in $U(1)$, has a massless phase,
but no long range order.  Lore is that higher symmetries only have
massive phases, but Seiler and Patrascioiu have argued that this may
be wrong.

\bigskip
I now turn to the renormalization group, which I approach via the
Migdal-Kadanoff approximate recursion relations.  I start with a
discussion of decimation.  Let us go back to the Ising model in one
dimension
$$
Z={\rm Tr}\pmatrix{
e^{\beta+H} & e^{-\beta}\cr
e^{-\beta} & e^{\beta-H}\cr
}^N={\rm Tr}T^N
$$
Let me sum over every other spin, giving
$$
Z={\rm Tr}(T^\prime)^{N/2}
$$
where
$$
T^\prime=T^2=
\pmatrix{
e^{2(\beta+H)}+e^{-2\beta} & e^{H}+e^{-H}\cr
e^{H}+e^{-H} & e^{2(\beta-H)}+e^{-2\beta}\cr
}
$$
We now match this with the original form of $T$
$$
T^\prime=C\pmatrix{
e^{\beta^\prime+H^\prime} & e^{-\beta^\prime}\cr
e^{-\beta^\prime} & e^{\beta^\prime-H^\prime}\cr
}
$$
We see that exactly the same physics occurs on a lattice of twice the
spacing and new couplings $(\beta^\prime, H^\prime)$.  The values of
$C$, $\beta^\prime$ and $H^\prime$ are fixed by the three equations
$$\eqalign{
C e^{-\beta^\prime}&=e^{H}+e^{-H}\cr
Ce^{\beta^\prime+H^\prime}&=e^{2(\beta+H)}+e^{-2\beta}\cr
Ce^{\beta^\prime-H^\prime}&=e^{2(\beta-H)}+e^{-2\beta}\cr
}
$$
This process is called decimation, {\it i.e.} integrating out some of the
degrees of freedom.  To simplify the equations, I turn off $H$,
obtaining
$$
\beta^\prime={1\over 2} \log(\cosh(2\beta)) 
$$
This can be written in a form reminiscent of our earlier recursions
$$
\beta^\prime-\beta=-{1\over 2}\log\left({2\over 1+e^{-4\beta}}\right) 
$$
The only fixed point occurs at $\beta=0$.  The new coupling is always
less than the old one as long as beta is positive.  Repeating this as
an iteration drives any $\beta$ to zero.

It is instructive to extend this to non integer decimations.  For
this, write the transfer matrix in the form
$T_{ss^\prime}=\cosh(\beta)(1+s s^\prime\tanh(\beta))$.  The above
decimation by a factor of two involves the sum
$$
\half \sum_{s_2} (1+s_1s_2t)(1+s_2s_3t)=(1+s_1s_3t^2)
$$
or simply
$\tanh(\beta)\rightarrow
\tanh^2(\beta)$.  Interpolate this to rescaling by a factor of
$1+\Delta$, taking $\tanh(\beta)
\rightarrow \tanh^{1+\Delta}(\beta)$.  Infinitesimally, this reduces to 
$$
{\beta^\prime-\beta\over\Delta}\sim a{d\beta\over
da}=\cosh(\beta)\sinh(\beta)
\log(\tanh(\beta))
$$
This is the renormalization group equation for this system.  The right
hand side is negative for all positive $\beta$.  As the lattice
spacing varies from zero to infinity, the coupling $\beta$ flows
from the ultraviolet fixed point at infinity to the infrared fixed
point at zero.

Suppose a system has a non-trivial fixed point satisfying
$$
a{d\beta\over da}=\lambda(\beta-\beta_c).
$$
This has the solution
$$
\beta=\beta_c+Ca^\lambda
$$
Since $a\sim 1/\xi$, this says
$$
1/\xi \sim (\beta-\beta_c)^{1/\lambda}
$$
This is the renormalization group way of seeing how non-trivial
exponents can arise as one approaches a critical point.

Going on to more dimensions we loose the exactness and must make
approximations.  Integrating out a site in more than one dimension
introduces couplings between all sites to which it is coupled.
Integrating the sites along a line couples all spins attached to that
line.  Integrating out all but the corners on a block requires an
infinite number of couplings.  This makes things less than rigorous,
but can imagine a similar coupling ``flow'' in a higher space.

\topinsert
\epsfxsize .7\hsize
\centerline{\epsffile{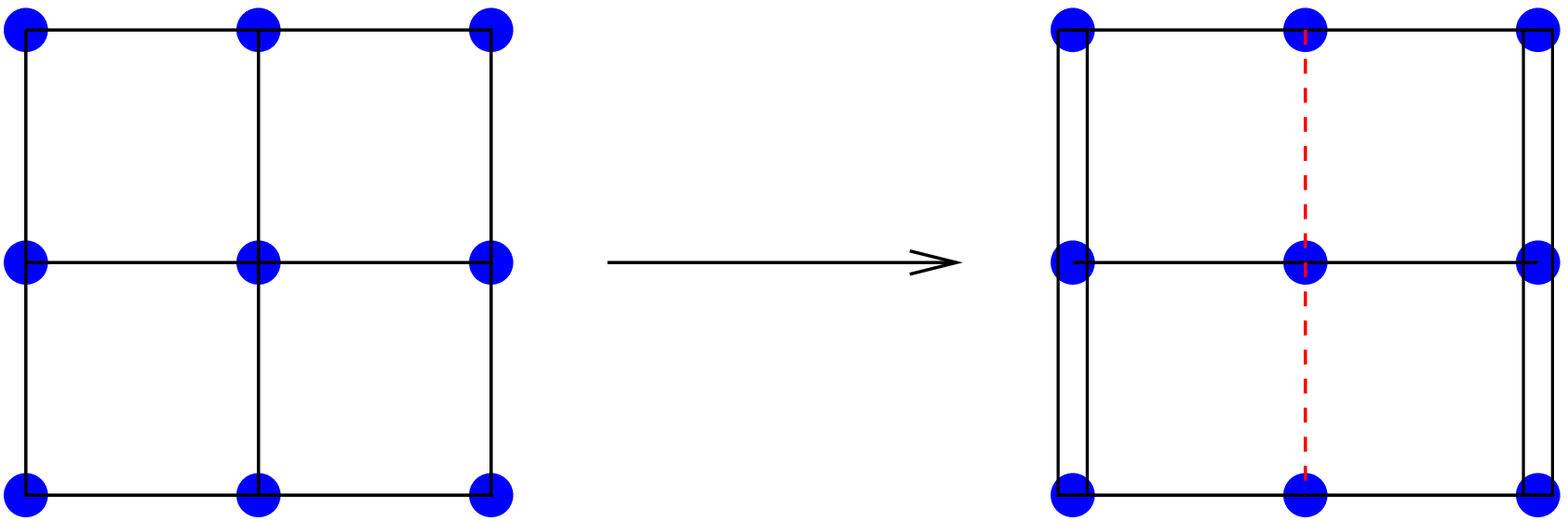}}
\centerline{Figure 9. Moving bonds around.}
\endinsert

Making an approximation by moving bonds around allows one to
analytically study these flows.  Imagine integrating out every other
site in say the $x$ direction.  To avoid long range couplings being
generated in the $y$ direction, follow Kadanoff and ``move'' the $y$
bonds to sites not being integrated over, as sketched in Fig.~9.
Every second $y$ bond becomes twice as strong, and then the earlier
$x$ decimation can be carried out on the remaining sites.  Thus we
relate the model at $\beta_x,\beta_y$ to that at
$$\eqalign{
\beta^\prime_x&={1\over 2} \log(\cosh(2\beta_x))\cr
\beta^\prime_y&=2\beta_y\cr
}
$$
Now repeat this for the $y$ direction.  The resulting transformation
is asymmetric due to the approximations.  To get more symmetric, do
things differentially, using the earlier equation for the $x$ coupling
and $a {d\beta\over da}=\beta$ for the bond moving.  The total change
of coupling is then
$$
a{d\beta\over
da}=\cosh(\beta)\sinh(\beta)
\log(\tanh(\beta))+(d-1)\beta
$$
Here I insert a factor of $d-1$ to allow for bond moving in all
directions but the decimation one.  The result is exact in one
dimension, and for $d=2$ it still gives the exact $\beta_c$.
I plot this function in Fig.~10 for $d=2$.

\topinsert
\epsfxsize .6\hsize
\centerline{\epsffile{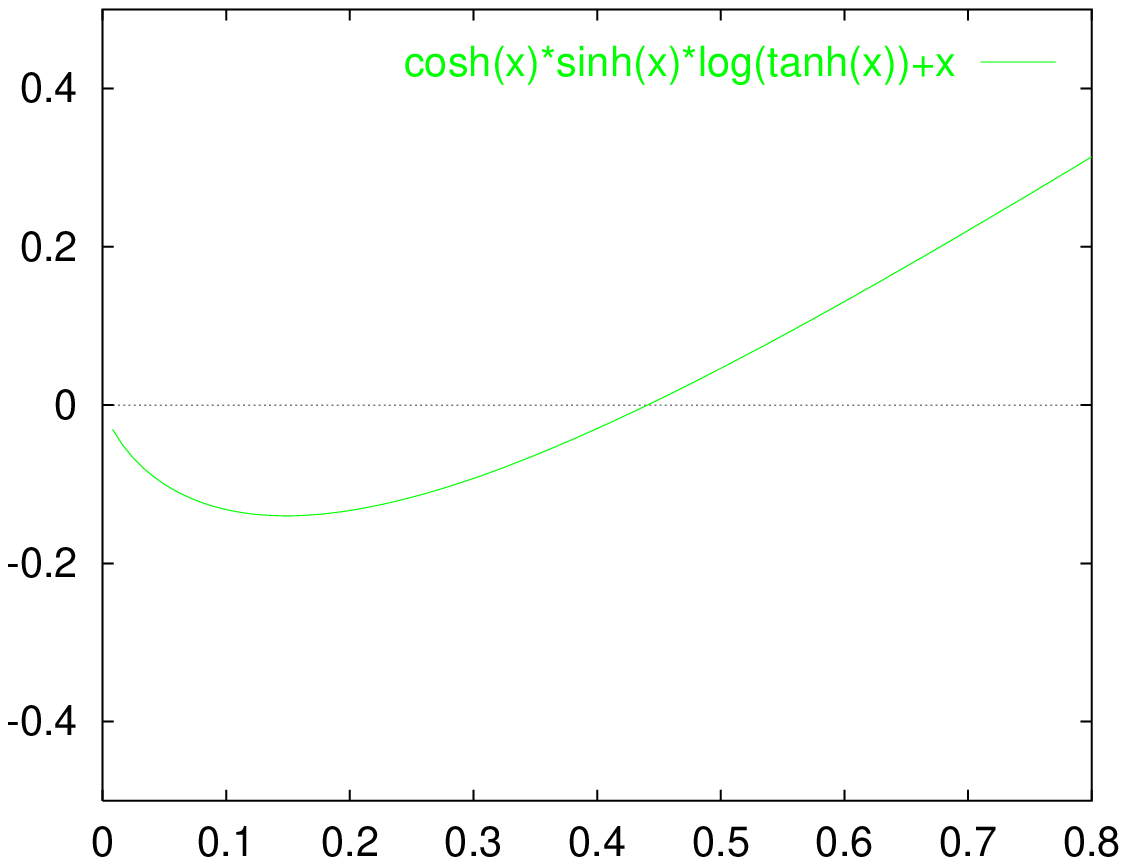}}
\centerline{Figure 10.  The Migdal-Kadanoff recursion relation for $d=2$.}
\endinsert

The renormalization group relates theories with different lattice
spacings.  If we could keep track of an infinite number of couplings,
the procedure would be ``exact,'' but in reality we usually need some
truncation.  Continuing to integrate out degrees of freedom, the
couplings flow and might reach some ``fixed point.''  With two
couplings, there can be an attractive ``sheet'' towards which
couplings flow, and then they go towards the fixed point, as sketched
in Fig.~11.  If the fixed point has only one attractive direction,
then two different models that flow towards that same fixed point will
have the same physics.  This is universality, {\it i.e.} exponents are
the same for all these models with the same attractor.

\topinsert
\epsfxsize .5\hsize
\centerline{\epsffile{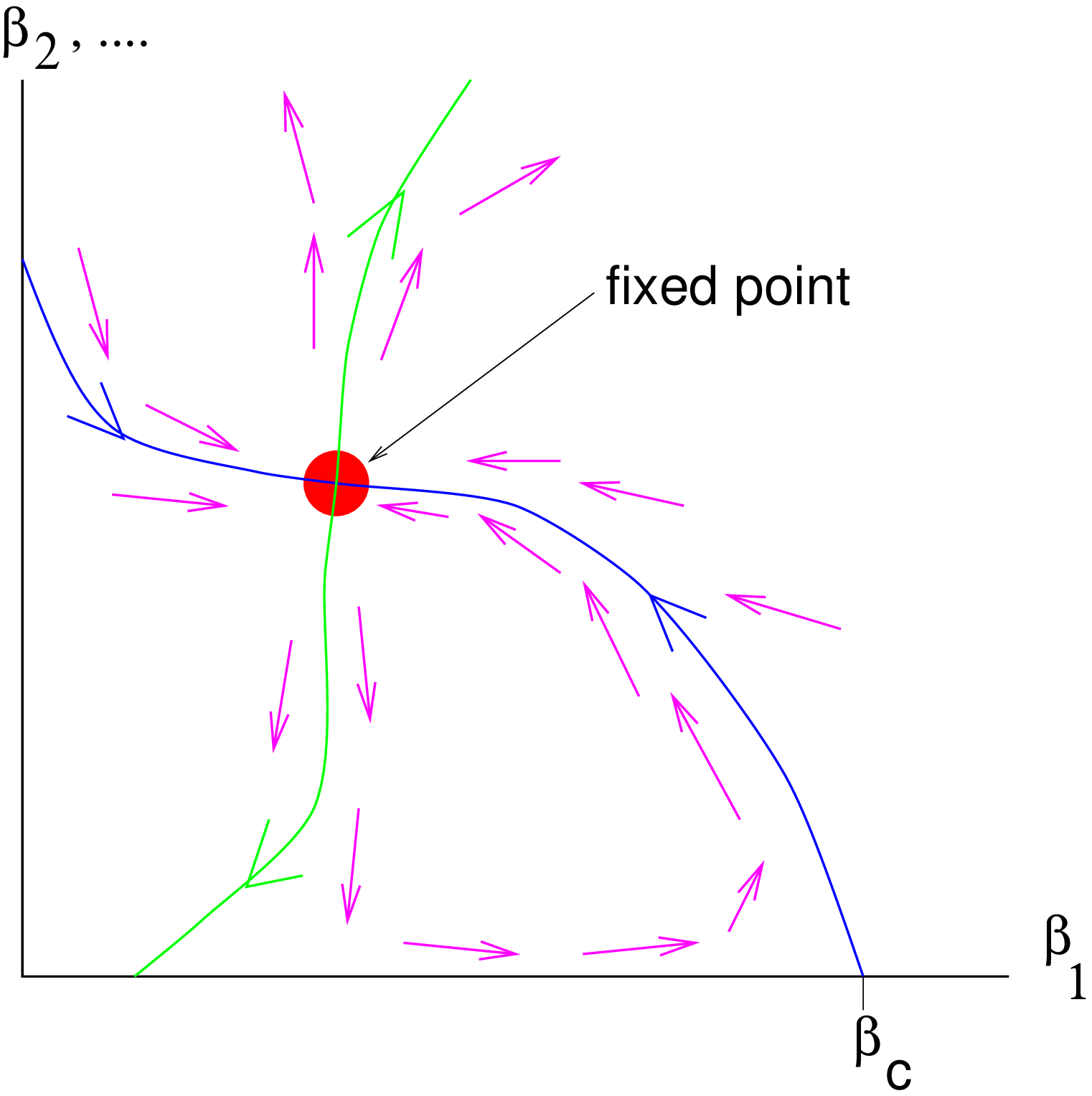}}
\centerline{Figure 11. A generic renormalization group flow.}
\endinsert

So mean field theory describes a phase transition in terms of a
changing classical ground state as parameters are varied, and the
renormalization group description is in terms of a flow through a
complex coupling constant space.  When should we rely on which
picture?

Some hints come from dimensional analysis, although, in ignoring
non-perturbative effects that might occur at strong coupling, the
following arguments are not rigorous.  In $d$ dimensions a
conventional scalar field has dimensions of $M^{d-2\over 2}$.  Thus
the coupling constant $\lambda$ in an interaction of form $\int d^dx\
\lambda \phi^n$ has dimensions of $M^{d-n{d-2\over2}}$.  On a lattice
of spacing $a$, the natural unit of dimension $M$ is the inverse
lattice spacing.  Thus without any special tuning, the renormalized
coupling at some fixed physical scale would naturally run as
$\lambda\sim a^{n{d-2\over2}-d}$.  As long as the exponent in this
expression is positive, i.e.
$$
n\ge {2d\over d-2}
$$ 
we expect the coupling to become ``irrelevant'' in the continuum
limit.  The fixed point is driven towards zero in the corresponding
direction.  If $d$ exceeds four, this is the case for all
interactions.  (I ignore $\phi^3$ in 6 dimensions because of stability
problems.)  This suggests that four dimensions is a critical case,
with mean field theory giving the right qualitative critical behavior
for all larger dimensions.  In four dimensions we have several
possible ``renormalizable'' couplings which are dimensionless,
suggesting logarithmic corrections to the simple dimensional
arguments.  Indeed, four-dimensional non-abelian gauge theories should
display exactly such a logarithmic flow; this is asymptotic freedom.

This simple dimensional argument applied to the mass term suggests it
would flow towards infinity in all dimensions.  For a conventional
phase transition, something must be tuned to a critical point.  In
statistical mechanics this is the temperature.  In field theory
language we usually remap this onto a tuning of the mass term, saying
that the transition occurs as some scalar mass goes through zero.
This tuning of scalar mass terms required for a continuum limit seems
unnatural and is one of the unsatisfying features of the standard
model, driving particle physicists to try to unravel how the Higg's
mechanism really works.

Recently there has been considerable interest in statistical systems
that become critical without any tuning of parameters.  This
phenomenon of ``self-organized criticality'' may explain the fractal
structure of much of the world around us.  However field theoretical
applications of this concept remain elusive.

In non-Abelian gauge theories with massless fermions, chiral symmetry
protects the mass from renormalization, avoiding any special tuning.
Indeed, these models exhibit the amazing phenomenon of dimensional
transmutation: all dimensionless parameters in the continuum limit are
completely determined by the basic structure of the initial
Lagrangian, without any continuous parameters to tune.  In the limit
of vanishing pion mass, the rho to nucleon mass ratio should be
determined from first principles; it is the goal of lattice gauge
theory to calculate just such numbers.

As we go below four dimensions, this dimensional argument suggests
that several couplings can become ``relevant,'' requiring the
renormalization group picture of flow towards a non-trivial fixed
point.  Above two dimensions the finite number of renormalizable
couplings corresponds to the renormalization group argument for a
finite number of ``universality classes,'' corresponding to different
basic symmetries.

One might imagine dimensionality as being a continuously variable
parameter.  Then just below four dimensions a renormalizable coupling
becomes ``super-renormalizable'' and a new non-trivial fixed point
breaks away from vanishing coupling.  Near four dimensions this point
is at small coupling, forming the basis for an expansion in $4-d$.
This has become a major industry, making remarkably accurate
predictions for critical exponents in three dimensions.

\bigskip
Now I return to lattice gauge theory and discuss how pure glue at
finite temperature mimics a three state model and might be expected to
have a first order deconfining transition.  I lead into this with a
bit of group theory.  Consider some compact group $G$ with elements
$g$.  There exists a unique measure
$$
\int dg f(g)=\int dg f(g_1 g)=\int dg f(g g_1)
$$
where I normalize $\int dg = 1$.  (Non-compact groups might have
different normalizations for left and right.)  For example, with
$U(1)$ I can take $g=e^{i\theta}$ and then $\int dg=\int_0^{2\pi}
d\theta/{2\pi}$.  For $SU(2)$, write $g=a_0+i\vec a \cdot \vec \sigma$
and then $\int dg=\int d^4 a\ \delta(a_0^2+\vec a^2-1)$, {\it i.e.}
the surface of a four dimensional sphere, an $S_3$.

This group integration extracts the ``singlet'' part of a function in
the following sense.  Suppose $f$ is a ``class function,''
{\it i.e.} $f(g)=f(g_1 g g_1^{-1})$.  Then I can expand it in traces over
the various irreducible representations $R$ of the group
$$
f(g)=\sum_R f_R \chi_R(g)
$$ 
where the character $\chi_R(g)={\rm Tr} M_R(g)$ and $M_R(g)$ is the
matrix representing $g$ in representation $R$.  These representations
include the trivial one, $R=0$, the fundamental one $R=F$, the adjoint
$R=A$, and generally infinitely many more.  For irreducible
representations, the characters $\chi$ satisfy an orthogonality
condition
$$
\int dg\ \chi^*_R(g)\ \chi_{R^\prime}(g)=\delta_{RR^\prime}
$$
From this and $\chi_0(g)=1$, we see
$$
\int dg f(g)=f_0.
$$
If we insert a character in some other representation
we obtain
$$
\int dg \chi^*_R(g)\ f(g)=f_R.
$$
This allows us to do some specific integrals integrals quite easily.
For example, with $SU(3)$ we have
$$
\int_{SU(3)} dg\ ({\rm Tr}g)^3=1
$$
since there is only one singlet in the famous decomposition $3\otimes
3\otimes 3= 1\oplus 8 \oplus 8 \oplus 10$.  This integral lies at the
base of the argument below for a first order chiral transition with
three massless quark flavors.

So lets apply this to lattice gauge theory.  On each bond of our
hyper-cubic lattice we have a group element $U_{ij}$.  The Wilson
action multiplies these around elementary squares and constructs
$U_P=U_1U_2U_3U_4$.  The partition function is
$$
Z=\int \{dU\} e^{-\beta\sum_P {\rm Re}\chi_F(U_P)}
$$
Formally, as argued earlier, this represents something like $Z={\rm
Tr}e^{a H N_t}$ with $a$ the temporal lattice spacing.  The picture,
however, is a bit more complicated due to gauge invariance.  If we put
a group element $g_i$ on each site, we can imagine taking
$U_{ij}\rightarrow g_i U_{ij} g_j^{-1}$.  This change cancels from the
action.

Gauge invariance leads to the possibility of gauge fixing.  This can
be done much more generally, but for now suppose I forget to integrate
over one link and define
$$
Z(U_0)=\int \{dU\}\ \exp(-\beta\sum_P \chi_0(U_P))\ \delta(U_{ij},U_0)
$$
On the integrand I can do a gauge transformation and then use the
invariance of measure to find
$$
Z(U_0)=Z(g_i U_0 g_j^{-1})
$$
So $Z(U_0)$ doesn't depend on $U_0$.  Since my measure is normalized,
$Z=Z_0$.  I can continue this and forget to integrate over more links.
As long as no closed loops are fixed, then the partition function is
unchanged.  A closed loop is a gauge invariant observable, so we
better not be able to fix it.

In the temporal gauge, all time-like links are set to unity.  In this
gauge the above transfer matrix argument reduces the path integral to
Hamiltonian lattice gauge theory.  But on finite periodic temporal
lattice, temporal links at a given spatial site form a closed loop.
Thus one cannot gauge fix all of them.  At each spatial location we
must leave one temporal link unfixed, take it to be at time 0.  What
does integrating over this link correspond to?

In Hamiltonian language, there is an operator $R_i(g)$ that does a
local gauge rotation at site $i$.  In particular, for a link to 
a spatial neighbor
$$
R_i(g) U_{ij} R_i^{-1}(g)=gU_{ij}
$$
These are all operators in the Hilbert space of the Hamiltonian
approach.  What the path integral formally reduces to is
$$
Z={\rm Tr} \left(e^{-aHN_t} \prod_i (\int dg_i R_i(g_i))\right)
$$
From the above discussion we see that this last integration projects
out a gauge singlet.  This operator imposes the 
lattice-gauge-theory version of Gauss's law
$$
\delta(\vec D \cdot \vec E)\leftrightarrow \int dg R_i(g)
$$

Now we can generalize and consider not projecting out the singlet
everywhere.  In particular, at one site I might want to put down a
quark-like source.  To do this I simply insert the character for the
desired representation
$$
\int dg \chi_F^*(g) R_i(g)
$$
The ratio of the new partition function to the old is the Wilson line
or the Polyakov loop.  Going back to the path integral, it is just the
expectation of a product of temporal links wrapping around the time
direction.  This Hamiltonian argument explicitly shows how it
represents the energy carried by a fixed source in the fundamental
representation of the gauge group.

For the quark-less theory, the unfixed temporal links at time 0 have a
global symmetry under the center of the gauge group.  For the $SU(3)$
of the strong interactions the center is the set $\{1,e^{\pm2\pi
i/3}\}$.  By definition, center elements commute with all group
elements, and the global change $g_i\rightarrow e^{2\pi i/3} g_i$ will
cancel out of the temporal plaquettes, each of which involves one
$g_i$ and one $g_j^{-1}$.  This is exactly the same symmetry as for
the 3 state Potts model, which I argued above should have a first
order phase transition.  This prediction is well verified by numerical
simulation.

The quark kinetic term explicitly breaks this symmetry, so the
transition might go away.  For massless quarks, it is instead the
global chiral symmetry that becomes relevant.  This suggests second
order for two flavors.  For three light flavors the suggestion is
first order since there is a quadratic term in the mean field
expansion.  This arises since the product of three fundamental
representations contains a singlet piece, as we well know because
three quarks can combine to form a gauge singlet baryon.  Generalizing
our earlier mean field equation to an $SU(3)$ spin system gives
$$
3M={\int dg\ e^{2d\beta M{\rm Re Tr} g}{\rm Tr} g \over 
\int dg\ e^{2d\beta M{\rm Re Tr} g}}
={d\beta M+(d\beta M)^2/2+ \ldots \over 1+(d\beta M)^2\ldots}
=d\beta M+(d\beta M)^2/2 \ldots 
$$
where I use the earlier $SU(3)$ example integral.  The quadratic term
means that the solution jumps discontinuously, just as argued earlier
for the Potts model.  The interpolation between the small and large
mass limits for various numbers of flavors is a major area of current
study.

\vfill\eject
\parindent=0pt
\parskip=6pt
\centerline{\bf A few general references}
\bigskip

P. Bak and M. Creutz, 1994, ``Fractals and self-organized
criticality,'' in {\sl Fractals in science}, Armin Bunde and Shlomo
Havlin eds., p. 26 (Springer-Verlag).  A review of self-organized
criticality.

Coleman, S., 1973, {\sl Commun. Math. Phys.} {\bf 31}, 259.  A
discussion of Goldstone Bosons in two dimensions.

Coleman, S., 1975, in {\sl Laws of hadronic matter}, A. Zichichi, ed.,
p. 139 (Academic Press).  An excellent discussion of effective
potentials and spontaneous symmetry breaking.

Coleman, S. and E. Weinberg, 1973, {\sl Phys. Rev.} {\bf D16},
1888.  Introduces dimensional transmutation.

Creutz, M., 1983, {\sl Quarks, gluons and lattices} (Cambridge
University Press).  An introduction to lattice gauge theory.

Itzykson, C. and J.-M. Drouffe, 1989, {\sl Statistical field theory},
(Cambridge University Press).  A nice general discussion of many of
the topics in these notes, especially duality, Cayley trees, and mean
field theory.

Kadanoff, L. 1976 {\sl Ann. Phys.} {\bf 100}, 359; {\sl Rev. Mod. Phys.}
{\bf 49}, 267; 
Migdal, A. 1975, {\sl Zh. Eksp. Teor. Fiz.} {\bf 69}, 810,1457 [{\sl
Sov. Phys. Jetp.} {\bf 42}, 413, 743].  The Migdal-Kadanoff recursion
relations.

Patrascioiu, A. and E. Seiler, 1997, hep-lat/9706011.  One of a series
of papers questioning common but unproven lore on two dimensional
models.

Shultz, T., D. Mattis, E. Lieb, 1964, {\sl Rev. Mod. Phys.} 36, 856.
An elegant solution of the two dimensional Ising model in terms of
domain walls being free fermions.

Wilson, K., (1971), {\sl Phys. Rev.} B4, 3174, 3184.  The basic
renormalization group idea as applied to critical phenomena.

\bye